\documentclass[%
 twocolumn,
 amsmath,amssymb,
 aps,
 pra,
]{revtex4-1}

\usepackage{amsmath,amssymb}
\usepackage[english]{babel}
\usepackage{ bbold }
\usepackage{upgreek}
\usepackage{dsfont}
\usepackage{graphicx}
\usepackage{graphicx}

\newcommand{\op}[1]{{\sf #1}}

\DeclareMathOperator\erf{erf}

\begin{document}
\title{Collapse-induced Orientational Localization of Rigid Rotors}

\author{Bj\"{o}rn Schrinski}
\author{Benjamin A. Stickler}
\author{Klaus Hornberger}
\affiliation{
 University of Duisburg-Essen, Faculty of Physics, Lotharstra\ss e 1, 47048 Duisburg, Germany}

\begin{abstract}
We show how the ro-translational motion of anisotropic particles is affected by the model of Continuous Spontaneous Localization (CSL), the most prominent hypothetical modification of the Schr\"odinger equation restoring realism on the macroscale.
We derive the master equation describing collapse-induced spatio-orientational decoherence, and demonstrate how it leads to linear- and angular-momentum diffusion. Since the associated heating rates scale differently with the CSL parameters, the latter can be determined individually by measuring the random motion of a single levitated nanorotor.
\end{abstract}

\maketitle

\section{Introduction} 
It is an open question whether the quantum superposition principle holds at arbitrary length and mass scales. Its validity has been confirmed 
for increasingly large particles and distances by interference experiments involving single neutrons and atoms \cite{zawisky2002neutron,AtomInterfero1,AtomInterfero2,AtomInterfero3},  Bose-Einstein condensates \cite{Shin2004,Jo2007,Baumgaertner2010,Schmiedmayer2013,Schmiedmayer2014,Kasevich}, and massive molecules \cite{MolInterfero1,MolInterfero2,MolInterfero4}. By direct verification of quantum superpositions \cite{KlausMarkusNature}, these experiments establish our perception of the quantum-to-classical transition.

Spontaneous localization models \cite{CSLReview,bassi2003dynamical} and the gravitational localization discussed by Di\'osi and Penrose \cite{diosi1984gravitation,Penrose1996} provide a dynamical description of the wavefunction collapse through an objective modification of the Schr\"odinger equation. One of the most prominent collapse theories is the continuous spontaneous localization (CSL) model \cite{Ghirardi1986CSL,Pearle1989CSL,Ghirardi1990CSL}. By adding a nonlinear and stochastic term to the Schr\"odinger equation it describes the random localization of the wave function of a delocalized particle. The modification involves two parameters, the rate $\lambda_{\rm C}$ and the localization length $r_{\rm C}$.  The maximal decay rate $\lambda_{\rm C} M^2/m_0^2$  grows quadratically with particle mass $M$ (in units of the reference mass $m_0$); it is met if the spatial delocalization well exceeds the CSL length $r_{\rm C}$, while spatial superpositions much smaller than $r_{\rm C}$ remain 
practically undisturbed. In this way the CSL model gives rise to classical behavior of massive particles, while the quantum dynamics of microscopic particles is preserved.

A natural way to test objective collapse models is to observe matter-wave interference and thus exclude all parameters $\lambda_{\rm C}$ and $r_{\rm C}$ that contradict the experiment \cite{Nimmrichter2011}. However, the CSL modification implies that even a localized particle isolated from its environment experiences momentum diffusion and heating, which can be measured in principle \cite{Bahrami2014heating,nimmrichter2014optomechanical,Diosi2014heating,Laloe2014BECheating,goldwater2015testing,li2016discriminating}. In fact, the most stringent up-to-date restrictions on the CSL parameters result from the noise spectrum of LISA pathfinder \cite{Carlesso2016LISA} and from X-ray emission from a Germanium surface \cite{curceanu1502x}.

Levitated nanoparticles offer a particularly promising platform for testing collapse models because they can  be well isolated from their environment \cite{optomechanicsReview1,chang2010cavity,romero2010toward,li2011millikelvin,Gieseler2012Nanospheres,romero2012quantum,kiesel2013Nanospheres,asenbaum2013cavity,optomechanicsReview2,Millen2015Nanospheres,Vovrosh2016,Rusconi2016magnetic}. So far, studies and tests of CSL involve only the center-of-mass degrees of freedom of the nanoparticle \cite{goldwater2015testing,li2016discriminating}. Motivated by recent advances in optical trapping and controlling the orientation of anisotropic nanoparticles \cite{Kuhn2015,hoang2016,kuhn2016} and the prospect to achieve ro-translational ground state cooling \cite{stickler2016,Zhong2017}, we study the effect of the CSL model on the orientational coherences of an arbitrarily shaped rigid object. 

Specifically, we present the master equation describing the impact of CSL on the ro-translational quantum dynamics of a rigid nanoparticle. The associated localization rate for orientational superpositions is then evaluated for bodies with cylindrical and spheroidal shape.
We determine the resulting linear momentum and angular momentum diffusion constants quantifying the rate of motional heating. Their dependence on the CSL parameters implies that the anisotropic shape of a nanoparticle can be exploited in future CSL tests. Finally, by solving the master equation for planar rotations we show how CSL leads to an enhanced spread of the orientational wave packet dynamics.

\section{Spatio-orientational localization}

All observable effects of CSL are accounted for by considering the modified time evolution of the state operator, $\partial_t\rho=-i[\op{H},\rho]/\hbar+\mathcal{L}\rho$. The modification $\mathcal{L}\rho$ to the von Neumann equation can be cast in Lindblad form as \cite{bassi2003dynamical}
\begin{equation} \label{eq:csl1}
\mathcal{L}\rho=-\frac{\lambda_{\rm C}}{2r_{\rm C}^3\pi^{3/2}m_0^2}
\int d^3\textbf{s}\hspace{1mm}\left[\op{M}(\textbf{s}),\left[\op{M}(\textbf{s}),\rho\right]\right].
\end{equation}
The mass operators $\op{M}(\textbf{s})$ depend on the position operators $\op{r}_n$ and masses $m_n$ of the constituent atoms,
\begin{equation}
\op{M}(\textbf{s})=\sum_n m_n e^{-(\textbf{s}-\op{r}_n)^2/2r^2_{\rm C}}.
\end{equation}

\subsection{Master equation for a rigid body}

The  position of the individual atoms can be specified by the center-of-mass position $\textbf{R}_{\rm cm}$, the orientation $\Omega$  (parameterized e.g.\ with Euler angles) and the displacements $\Delta \textbf{r}_n$ from the orientation-dependent equilibrium position $\op{R}({\Omega})\textbf{r}_n^{(0)}$, i.e., $\textbf{r}_n= \textbf{R}_{\rm cm} + \op{R}({\Omega})\textbf{r}_n^{(0)}+\Delta\textbf{r}_n$. Here, the matrix $\op{R}(\Omega)$ rotates from the body-fixed frame to the space-fixed frame $\{\textbf{e}_1,\textbf{e}_2,\textbf{e}_3\}$. 
If the displacements $\Delta\textbf{r}_n$ of the individual atoms around their equilibrium position are small compared to $r_{\rm C}$ it is admissible to describe the particle as a rigid body with position $\textbf{R}_{\rm cm}$ and orientation $\Omega$.

Denoting the mass density by $\varrho[\op{R}^T({\Omega})\textbf{r}]= \sum_n m_n \delta[\op{R}^T({\Omega})\textbf{r}-\textbf{r}_n^{(0)}]$, and the position and orientation operators as $\op{R}_{\rm cm}$ and $\op{\Omega}$, one can rewrite the mass operator $\op{M}(\textbf{s})$ as
\begin{align} \label{eq:lindbl}
\op{M}(\textbf{s})&\approx\int d^3\textbf{r}\,
\varrho[\op{R}^T(\op{\Omega})\textbf{r}] e^{-(\textbf{s}-\op{R}_{\rm cm} - \textbf{r})^2/2r^2_C}\nonumber\\
&=\frac{r^3_C}{(2\pi)^{3/2}}\int d^3 \textbf{k}\,
e^{-r^2_C k^2/2+i\textbf{k}\cdot(\textbf{s}-\op{R}_{\rm cm})}\tilde{\varrho}[\op{R}^T(\op{\Omega})\textbf{k}].
\end{align}
The Fourier transform of the mass density $\tilde{\varrho}(\textbf{k})$, often referred to as form factor, is normalized to the total mass of the particle, $\tilde\varrho(0) = M$.

The ro-translational master equation is obtained by inserting the mass operators (\ref{eq:lindbl}) into Eq.~(\ref{eq:csl1}). Its incoherent part takes the form
\begin{align} \label{eq:csl2}
\mathcal{L}\rho=&\frac{r^3_{\rm C}\lambda_{\rm C}}{\pi^{3/2}m_0^2}\int d^3\textbf{k}\hspace{1mm}e^{-r^2_{\rm C}k^2}\nonumber\\
&\times\Bigg[e^{-i\textbf{k}\cdot\op{R}_{\rm cm}}\tilde{\varrho}[\op{R}^T(\op{\Omega})\textbf{k}]\rho\tilde{\varrho}^*[\op{R}^T(\op{\Omega})\textbf{k}]e^{i\textbf{k}\cdot\op{R}_{\rm cm}}\nonumber\\
&\phantom{\times\Big\{}-\frac{1}{2}\left\{ \left \vert\tilde{\varrho}[\op{R}^T(\op{\Omega})\textbf{k}] \right \vert^2,\rho\right\}\Bigg].
\end{align} 
 Being diagonal in position and orientation it describes spatio-orientational decoherence. Evaluating the matrix elements $\left<\textbf{R}_{\rm cm}\Omega\right|\mathcal{L}\rho\left|\textbf{R}_{\rm cm}'\Omega'\right>$ one finds that the decay of the spatial off-diagonal elements depends not only on the distance $|\textbf{R}_{\rm cm}-\textbf{R}_{\rm cm}'|$  but also on the direction of displacement. We will see next that the decay of orientational coherences depends only on the relative orientation $\widetilde \Omega(\Omega,\Omega')$ defined by $\op{R}(\widetilde \Omega)=\op{R}^T(\Omega') \op{R}(\Omega) . $ 

 \begin{figure} 
  \centering
  \includegraphics[width=0.48\textwidth]{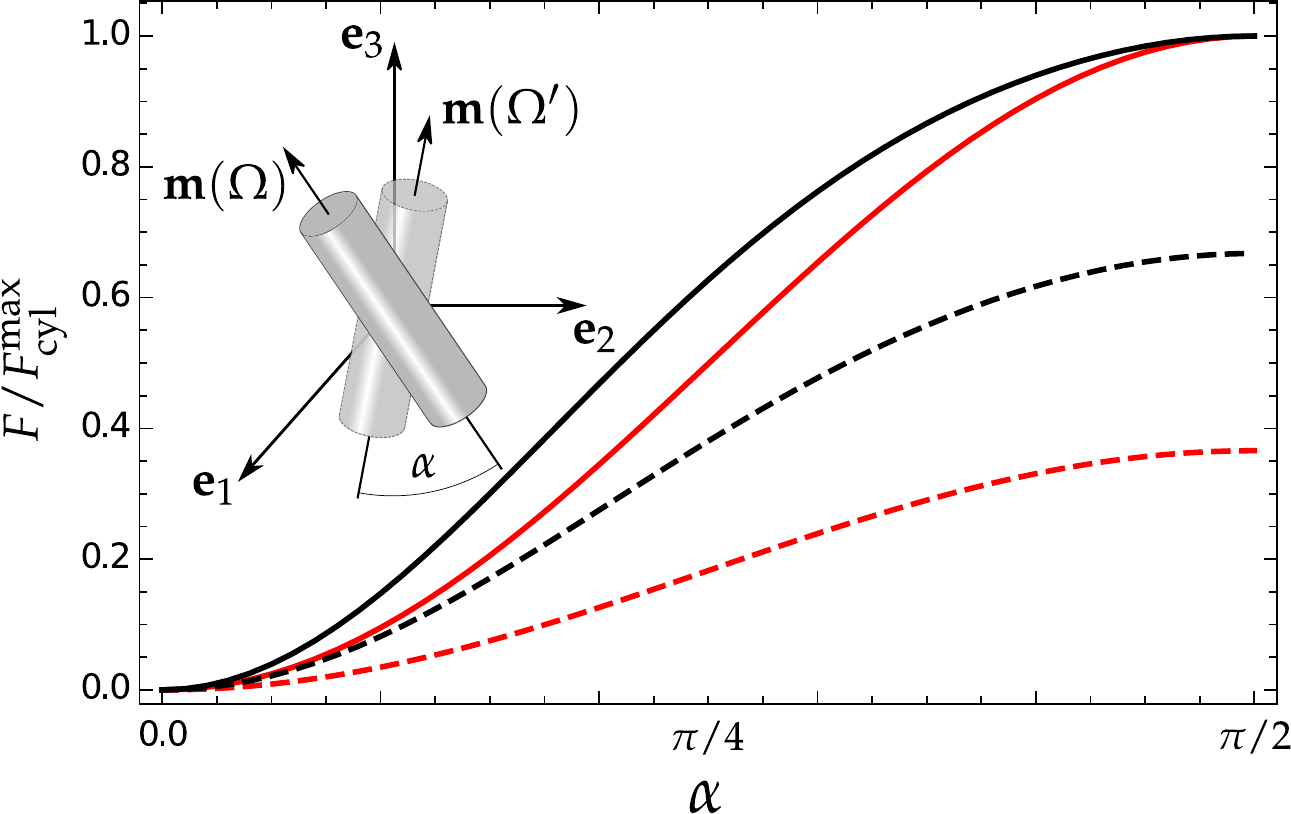}
  \caption{Orientational localization rate \eqref{eq:loc} for a cylinder (solid lines) and a spheroid (dashed lines) as a function of the angle $\alpha = {\rm arccos}[\textbf{m}(\Omega) \cdot \textbf{m}(\Omega')]$ in units of the maximal localization rate of the cylinder. For $ r_{\rm C}=L$ (red lines) the localization rate is proportional to $\vert \textbf{m}(\Omega) \times \textbf{m}(\Omega') \vert^2$. This simple $\sin^2 \alpha$-behavior is lost for $ r_{\rm C}=L/10$ (black lines). Throughout the figure we set $L / R = 20$, as motivated by \cite{Kuhn2015,kuhn2016}.}
\label{fig:LocalizationRate}
\end{figure}

\subsection{Orientational Localization} 
Tracing out the center-of-mass degrees of freedom in Eq.~(\ref{eq:csl2}) shows how pure orientational superpositions decay under the CSL modification. Specifically, the localization rate $F(\Omega,\Omega')$, defined as
\begin{align}
\left<\Omega\right|\mathrm{tr}_{\rm cm}(\mathcal{L} \rho)\left|\Omega'\right>
=&-F(\Omega,\Omega')
\left<\Omega\right|\mathrm{tr}_{\rm cm}(\rho)\left|\Omega'\right>,
\end{align}
is given by
\begin{align} \label{eq:loc}
F(\Omega,\Omega')=&\frac{r^3_{\rm C}\lambda_{\rm C}}{2\pi^{3/2}m_0^2}\int d^3\textbf{k}\hspace{1mm}e^{-r^2_{\rm C}k^2}\nonumber\\
&\hspace{17mm}\times\left\vert \tilde{\varrho}[\op{R}^T(\Omega)\textbf{k}]-\tilde{\varrho}[\op{R}^T(\Omega')\textbf{k}]\right\vert^2.
\end{align}
Its  imaginary part vanishes even for complex form factors, $\tilde{\varrho}(\textbf{k})=\tilde{\varrho}^*(-\textbf{k})$,  
due to the odd symmetry of the integrand.
As anticipated, the localization rate depends  on the relative orientation $\op{R}(\widetilde \Omega)$, as follows from rotating $\textbf{k}$.  

For azimuthally symmetric bodies the localization rate \eqref{eq:loc}  is only a function of the angle between the symmetry axes $\textbf{m}(\Omega)$ and $\textbf{m}(\Omega')$.
It is depicted in Fig. \ref{fig:LocalizationRate} for the case of a cylindrically and a spheroidally shaped homogeneous mass density of radius $R$ and length $L$. The respective form factors $\tilde\varrho(\textbf{k})$ are 
\begin{align}
\tilde{\varrho}_{\rm cyl}(\textbf{k})&=
\frac{2M}{R|\textbf{e}_3\times\textbf{k}|}J_1(R|\textbf{e}_3\times\textbf{k}|)\,{\rm sinc}\left(\frac{L}{2}\textbf{e}_3\cdot\textbf{k}\right),\\
\tilde{\varrho}_{\rm sph}(\textbf{k})&=
M\sqrt{\frac{9\pi}{2}}\frac{J_{3/2}(\sqrt{R^2|\textbf{e}_3\times\textbf{k}|^2+L^2|\textbf{e}_3\cdot\textbf{k}|^2/4})}{(\sqrt{R^2|\textbf{e}_3\times\textbf{k}|^2+L^2|\textbf{e}_3\cdot\textbf{k}|^2/4})^{3/2}},
\end{align}
where we chose the symmetry axis along $\textbf{e}_3$; $J_n(\cdot)$ are Bessel functions. 

\begin{figure*}
  \centering
  \includegraphics[width=1\textwidth]{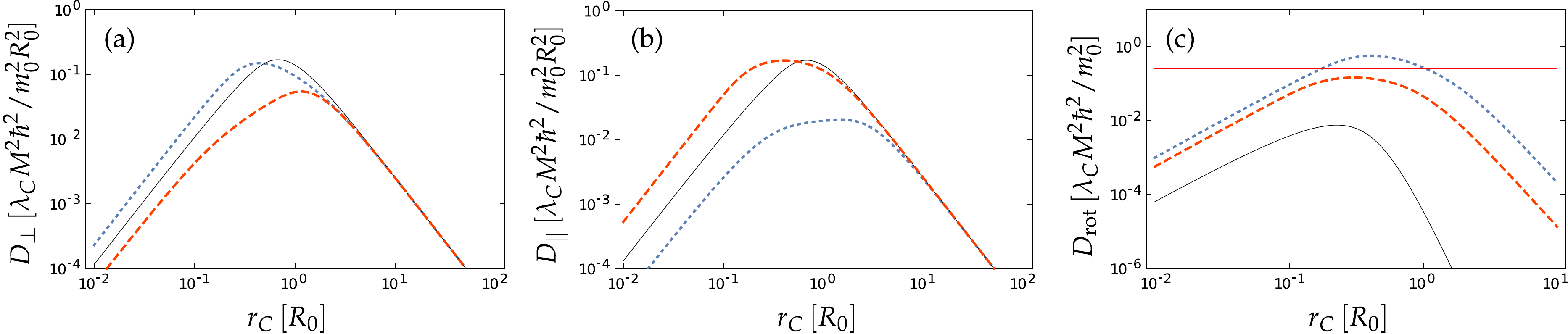}
  \caption{CSL induced diffusion constants $D_{\perp},D_{\parallel}$ and $D_{\rm rot}$ for a rigid cylinder as a function of the localization length $r_{\rm C}$. The black solid curve corresponds to a cylinder with minimal anisotropy, $L/R=\sqrt{3}$, the blue dotted lines to a rod with $L/R=8\sqrt{3}$, and the red dashed curves to a disc with $L/R=\sqrt{3}/8$. The volume is kept constant to ensure comparability, $V = \sqrt{3} \pi R_0^3$. (a),(b) The linear momentum diffusion coefficients depend strongly  on the shape of the nanoparticle and are maximal for diffusion perpendicular to the main extension of the nanoparticle. (c) The angular momentum diffusion coefficient is minimal for a cylinder with isotropic tensor of inertia, $L / R = \sqrt{3}$, and increases with increasing cylinder length $L$. In contrast, $D_{{\rm rot}}$ is bounded from above for discs (horizontal red line), as discussed in the text.}
\label{fig:DiffusionPlots}
\end{figure*}

If the maximum extension of the nanoparticle is well below the localization length $r_{\rm C}$, \eqref{eq:loc} takes on the form
\begin{align}
F(\Omega,\Omega')
\approx&\frac{\lambda_{\rm C} M^2}{8 m_0^2r^4_{\rm C}}\left(\frac{R^2}{a} - \frac{L^2}{b}\right)^2 \left \vert\textbf{m}(\Omega)\times\textbf{m}(\Omega') \right \vert^2,
\label{ThetaPhiIntegral}
\end{align}
where $a$, $b$ are numerical constants, $a_{\rm cyl}=4$, $b_{\rm cyl}=12$ and $a_{\rm sph}=5$, $b_{\rm sph}=20$. 

This rate increases with $\sin^2 \alpha= \vert\textbf{m}(\Omega)\times\textbf{m}(\Omega')  \vert^2$, as also observed for environmentally induced orientational decoherence of small particles \cite{stickler2016spatio,zhong2016}. Note that Eq.~(\ref{ThetaPhiIntegral}) vanishes if the spheroid is deformed into a sphere, $R=L/2$; the same holds for cylinders with isotropic tensor of inertia, i.e. $R=L/\sqrt{3}$. Remarkably, the localization rate (\ref{ThetaPhiIntegral}) scales with the \emph{tenth } power of the particle size (at fixed density). This holds for arbitrarily shaped small particles, as follows from dimensional analysis of \eqref{eq:loc}.

\section{Linear- and angular-momentum diffusion} 

The exact master equation (\ref{eq:csl2}) of spatio-orientational decoherence can be simplified for states which are sufficiently well localized in position and orientation (around $\Omega_0$) implying that $\left<\textbf{R}_{\rm cm}\Omega\right|\rho\left|\textbf{R}_{\rm cm}'\Omega'\right>\approx 0$ unless $|\textbf{R}_{\rm cm}-\textbf{R}_{\rm cm}' +\op{R}(\Omega_0) [\op{R}(\widetilde{\Omega})-\mathbb{1}] \textbf{r}_n^{(0)}|\ll r_{\rm C}$ for all $n$. This requirement demands that the relative  orientation is small, $\op{R}(\widetilde{\Omega}) \approx \mathbb{1} + \epsilon_{ijk}\,\mathrm{d}\Omega_i\,\textbf{e}_k\otimes\textbf{e}_j$  with $\mathrm{d}\Omega_i$ the angle of rotation around the $\textbf{e}_i$ axis.

Expanding the CSL modification $\mathcal{L} \rho$ into lowest order of $|\textbf{R}_{\rm cm}-\textbf{R}_{\rm cm}'|$ and $\mathrm{d} \Omega_i$ shows that the total localization is given by a sum of  pure center-of-mass and pure orientational localization,
\begin{align}
&\left<\textbf{R}_{\rm cm}\Omega\right|\mathcal{L}\rho\left|\textbf{R}_{\rm cm}'\Omega'\right>
\approx - \left<\textbf{R}_{\rm cm}\Omega\right|\rho\left|\textbf{R}_{\rm cm}'\Omega'\right> \nonumber\\
&\hspace{23mm}\times \left[F_{\rm cm}(\textbf{R}_{\rm cm} - \textbf{R}'_{\rm cm},\Omega_0)+F_{\rm rot}(\Omega,\Omega')\right].
\end{align}  
The respective rates can be expressed as
\begin{subequations}\label{eq:loc2}
\begin{align} 
F_{\rm cm}(\textbf{R},\Omega_0)=&
\frac{\lambda_{\rm C}}{2r_{\rm C}^2}\textbf{R} \cdot
\op{R}(\Omega_0) \op{A}_{\rm cm} \op{R}^T(\Omega_0)\textbf{R}\\
F_{\rm rot}(\Omega,\Omega')=&
\frac{\lambda_{\rm C}}{2}\mathrm{d}\boldsymbol{\Omega} \cdot
\op{A}_{\rm rot}\mathrm{d}\boldsymbol{\Omega},
\end{align}
\end{subequations}
where $\mathrm{d} \boldsymbol{\Omega} = (\mathrm{d} \Omega_1,\mathrm{d} \Omega_2,\mathrm{d} \Omega_3)$. Here the  geometry tensors
\begin{subequations} \label{eq:geomtens}
\begin{align}
\op{A}_{\rm cm}=&\frac{r^5_C}{\pi^{3/2}m_0^2}\int d^3\textbf{k}\hspace{1mm}e^{-k^2r_{\rm C}^2}|\tilde{\varrho}(\textbf{k})|^2
\textbf{k}\otimes\textbf{k}\,,
\\
\op{A}_{\rm rot}=&\frac{r^3_C}{\pi^{3/2}m_0^2}\int d^3\textbf{k}\hspace{1mm}e^{-k^2r_{\rm C}^2} [\textbf{k}\times\nabla_{\textbf{k}}\tilde{\varrho}(\textbf{k})]\otimes[\textbf{k}\times\nabla_{\textbf{k}}\tilde{\varrho}(\textbf{k})]\,,
\end{align}
\end{subequations}
account for the nonspherical shape of the matter density. They
naturally generalize the geometry factor for momentum diffusion in a single spatial direction as defined in Ref. \cite{nimmrichter2014optomechanical}.

\subsection{Azimuthally symmetric bodies}

The tensors $\op{A}_{\rm cm}$ and $\op{A}_{\rm rot}$ share the symmetries of the mass density $\varrho(\textbf{r})$.  In particular, for an azimuthally symmetric body also invariant under spatial inversion, like a cylinder or a spheroid, the geometry tensors have the general form
\begin{subequations}\label{LocParameterCylSymBody}
\begin{align}
\op{A}_{\rm cm}&= \frac{2 r_{\rm C}^2}{\lambda_{\rm C} \hbar^2} \left [D_{\perp}\mathbb{1}
+(D_{\parallel}-D_{\perp})\textbf{e}_{3}\otimes\textbf{e}_{3} \right ]\\
\op{A}_{\rm rot}&= \frac{2 D_{\rm rot}}{\lambda_{\rm C} \hbar^2} \left ( \textbf{e}_{1}\otimes\textbf{e}_{1} + \textbf{e}_{2}\otimes\textbf{e}_{2}\right ),
\end{align}
\end{subequations}
where we chose the symmetry axis to point into direction $\textbf{e}_3$. As demonstrated below, the constants $D_\|$, $D_\perp$ and $D_{\rm rot}$ are diffusion coefficients. For cylindrical bodies they are specified in the appendix. 

Plugging relations (\ref{LocParameterCylSymBody}) into (\ref{eq:loc2}) leads to the localization rates  
\begin{subequations} \label{eq:loc3}
\begin{align}
F_{\rm cm}({\bf R},\Omega_0) & = \frac{D_\bot}{\hbar^2} R^2 + \frac{ D_\| - D_\bot}{\hbar^2} \left [\textbf{R} \cdot \textbf{m}(\Omega_0)\right ]^2\\
F_{\rm rot}(\Omega,\Omega')& =\frac{D_{\rm rot}}{\hbar^2} \left | \textbf{m}(\Omega)\times\textbf{m}(\Omega') \right |^2.
\end{align}
\end{subequations}
The corresponding master equation reads
\begin{align}
&\partial_t  \rho =  -\frac{i}{\hbar} [\op{H},\rho] -\frac{D_\perp}{\hbar^2} \sum_{i = 1}^3 \left [ \op{R}_{i,\rm cm} , \left  [\op{R}_{i, \rm cm}, \rho \right ] \right] \notag \\
& - \frac{D_\| - D_\perp}{\hbar^2} \left [ \op{R}_{\rm cm} \cdot {\bf m}(\Omega_0),\left[\op{R}_{\rm cm} \cdot {\bf m}(\Omega_0),\rho \right ] \right ] \notag \\
& -  \frac{15 D_{\rm rot}}{4\hbar^2} \int_{S_2} \frac{d^2 \textbf{n}}{4 \pi} \left [  [\textbf{n} \cdot \textbf{m}(\op{\Omega})]^2 , \left [ [\textbf{n} \cdot \textbf{m}(\op{\Omega})]^2, \rho \right ] \right ]. 
 \label{eq:MeqDiffLim}
\end{align}
It describes linear- and angular momentum diffusion \cite{stickler2016spatio} as discussed next.

\subsection{Heating rates} 

In order to demonstrate that the master equation \eqref{eq:MeqDiffLim} indeed describes linear and angular momentum diffusion, we note that the expectation values of the linear momentum operator $\op{P}_{\rm cm}$ and of the angular momentum operator $\op{J}$ \cite{stickler2016spatio} are conserved,
\begin{subequations}
\begin{equation}
\partial_t \left<\op{P}_{\rm cm}\right> = 0, \qquad \qquad \partial_t \left<\op{J}\right> = 0,
\end{equation}
while the second moments increase linearly with time,
\begin{equation}
\partial_t \left<\op{P}_{\rm cm}^2\right> = 2 D_\| + 4 D_\perp, \qquad \qquad \partial_t \left<\op{J}^2\right> = 4 D_{\rm rot}.
\end{equation}
\end{subequations}
This follows with the canonical commutation relations by direct calculation. 
The linear- and angular-momentum heating rates are thus fully determined by the diffusion coefficients.

In Fig.~\ref{fig:DiffusionPlots} we show the diffusion coefficients of cylindrical bodies as a function of the localization length $r_{\rm C}$. Remarkably, angular momentum diffusion is stronger for long rods than for flat discs of the same volume, since the diffusion coefficient of disks is bounded, $D_{\rm rot} = \lambda_{\rm C} \hbar^2 M^2/4 m_0^2$ as $R / r_{\rm C} \to\infty$, while for long rods we find the asymptotic behavior $D_{\rm rot} \sim \sqrt{\pi} \lambda_{\rm C} L \hbar^2 M^2/24 r_{\rm C} m_0^2$ as $L / r_{\rm C} \to\infty$. We note that the diffusion constants of spheroidal particles agree qualitatively with those of cylinders.

\begin{figure}
  \centering
     \includegraphics[width=0.45\textwidth]{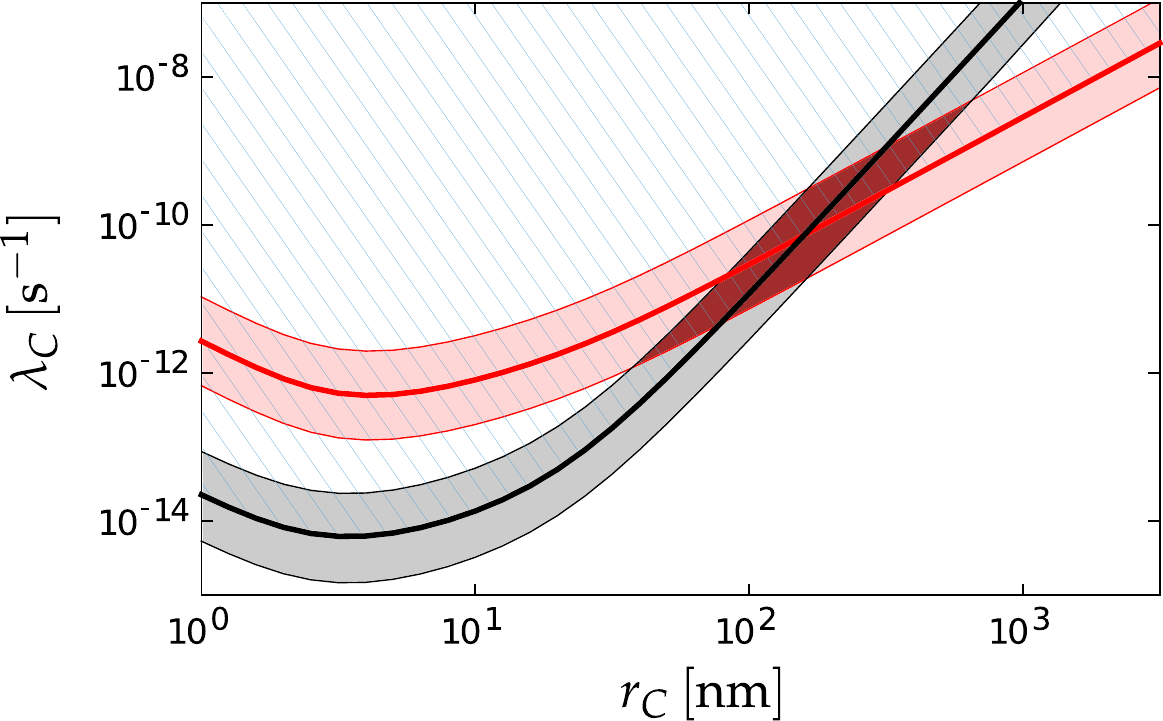}
  \caption{Measuring CSL parameters by simultaneous observation of center-of-mass and angular momentum diffusion: The red and black line represents the lower bound on the CSL localization rate found by a hypothetical measurement of the heating rates $\Gamma_{\rm cm}=   2 D_\perp/M = 10^{-8} {\rm K/s}$ and $\Gamma_{\rm rot}=  2D_{\rm rot}/I = 10^{-10} {\rm K/s}$ of a silicon spheroid with length $L = 100$\,nm and radius $R = 5$\,nm. This would exclude all CSL parameters from the blue hatched area. Provided all other sources of environmental decoherence are accounted for, the unassigned diffusion can be attributed to CSL.  The light red and light gray shaded areas indicate the measurement error associated with the inferred CSL rate.  The intersection of both lines allows one to extract the individual values of  $r_{\rm C}$ and $\lambda_{\rm C}$ in a single experiment, as shown by the dark 
red area. }
\label{CSLParameter}
\end{figure}

An important feature of the spatio-orientational localization, as compared to pure center-of -mass localization, is that the linear and the angular momentum diffusion coefficients depend differently on the CSL parameters $\lambda_{\rm C}$ and $r_{\rm C}$
even for small particles $L/r_{\rm C}\ll 1$.
Specifically, for small nanoparticles we have $D_{\rm rot} \propto \lambda_{\rm C} / r_{\rm C}^4$ while 
$D_{\rm cm}\propto \lambda_{\rm C} / r_{\rm C}^2$.
Thus,
simultaneously  measuring the linear- and angular-momentum diffusion coefficients of a levitated nanorotor would  allow one to determine both CSL parameters in a single experiment
(provided that environmental decoherence can be controlled). This is illustrated in Fig.~\ref{CSLParameter} where we show how the CSL loclization rate $\lambda_{\rm C}$ and the CSL length $r_{\rm C}$ would be extracted from a hypothetical measurement of the heating rates $\Gamma_{\rm cm}=   2 D_\perp/M = 10^{-8} {\rm K/s}$ and $\Gamma_{\rm rot}=  2D_{\rm rot}/I = 10^{-10} {\rm K/s}$ of a silicon spheroid.

\begin{figure*}
  \centering
  \includegraphics[width=1\textwidth]{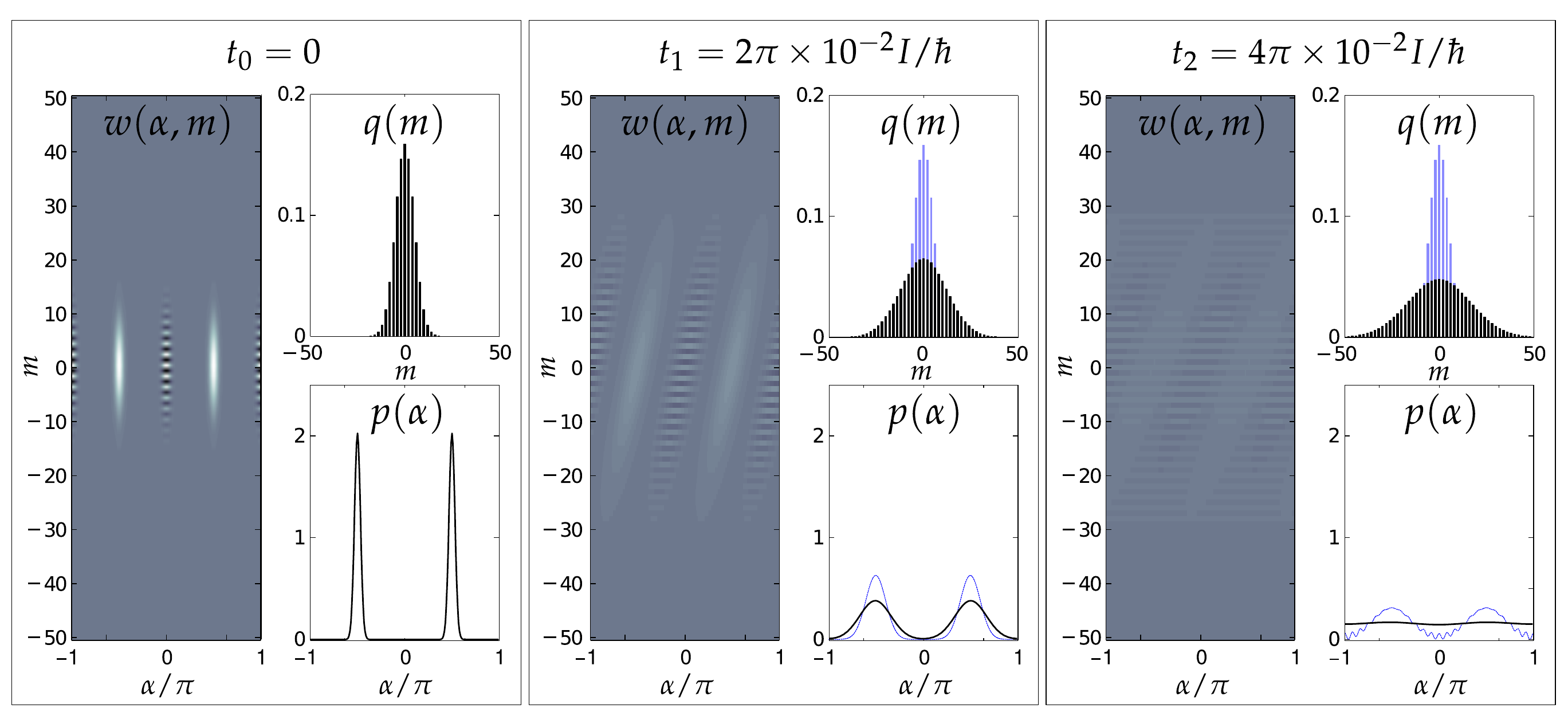}
  \caption{Time evolution (\ref{eq:PlanTE}) of the initial planar rotor state (\ref{eq:OrientGauss}) with width $\sigma_\alpha=0.1$ and CSL angular momentum diffusion constant $D_{\rm rot}=10^3\hbar^3/I$. We show the Wigner functions (density plot), and their marginals $p(\alpha)=\sum_m w(\alpha,m;t)$ and $p(m)=\int d\alpha\, w(\alpha,m;t)$  (black lines) for the times $t_0=0,$ $ t_1=2\pi\times 10^{-2} I/\hbar$, and $ t_2=4\pi\times 10^{-2} I/\hbar$. For comparison, the blue lines depict the marginal distributions without CSL modification.}
\label{fig:PlanarTE}
\end{figure*}

\section{Planar Rotations} 

As a simple and exactly solvable application, we study the impact of the CSL modification on the rotational dynamics of the planar rotor with inversion symmetry and a single orientational degree of freedom $\alpha \in [-\pi,\pi)$. The respective master equation can be obtained from \eqref{eq:MeqDiffLim} by tracing out the center-of-mass degrees of freedom and restricting the rotations to the $\textbf{e}_1$-$\textbf{e}_2$ plane,
\begin{align} \label{eq:planrot}
\partial_t\rho_{\rm rot}=&-\frac{i}{\hbar}\left[\frac{\op{p}_{\alpha}^2}{2I},\rho_{\rm rot}\right]\nonumber\\
&+\frac{2D_{\rm rot}}{\hbar^2\pi}\int_0^{2 \pi} d\phi\,\Big[
\cos^2(\phi-\upalpha)\rho_{\rm rot}\cos^2(\phi-\upalpha)\nonumber\\
&-\frac{1}{2}\left\{\cos^4(\phi-\upalpha),\rho_{\rm rot}\right\}
\Big].
\end{align}
Here, $I$ denotes the moment of ineratia and $\op{p}_\alpha=\op{J}\cdot\textbf{e}_3$.

The master equation \eqref{eq:planrot} can be conveniently expressed in phase space by defining the Wigner function for the orientation state \cite{Leonardt1995Wigner,Rigas2011Wigner}, 
\begin{equation} \label{eq:wignerf}
w(\alpha,m) = \int_{-\pi}^\pi \frac{d\alpha'}{2 \pi} e^{i m  \alpha'} \left<\alpha - \alpha'/2\right|\rho\left|\alpha + \alpha'/2\right>,
\end{equation}
where the angular momentum numbers $m$ are restricted to discrete values, $m \in \mathds{Z}$. This Wigner function, as a  special case of the Wigner function for general orientation states \cite{fischer2013Wigner}, is real, normalized, and gives the correct marginals. The resulting quantum Liouville equation takes on the form
\begin{align}
\partial_t w(\alpha&,m)+\frac{\hbar m}{I}\partial_\alpha w(\alpha,m)\nonumber\\
=&D_{\rm rot}\frac{w(\alpha,m-2)-2w(\alpha,m)+w(\alpha,m+2)}{(2\hbar)^2}\,.
\label{eq:diffrotplanar}
\end{align}
The CSL modification thus enters in the discretized form of a second order angular momentum derivative. 
The fact that only next-to-nearest angular momentum quantum number $m$ are coupled is due to the inversion symmetry of the rotor.

The solution of Eq.~(\ref{eq:diffrotplanar}) with the initial condition $w_0(\alpha,m)$ can be explicitly given as
\begin{align}
w(\alpha,m;t)
=\sum_{\ell\in \mathds{Z}}\int_{-\pi}^\pi d\alpha'\,w_0\left(\alpha-\alpha'-\frac{\hbar m t}{I},m-2\ell\right)T_t(\alpha',\ell)\,.
\label{eq:PlanTE}
\end{align}
The kernel
\begin{align}
T_t(\alpha',\ell)=&\frac{e^{-D_{\rm rot}t/2\hbar^2}}{2\pi}\sum_{k\in \mathds{Z}}
e^{ik(\alpha'+\ell\hbar t/I)}I_\ell\left[\frac{D_{\rm rot}t}{2\hbar^2}{\rm sinc}\left(\frac{\hbar k t}{I}\right)\right],
\label{eq:PlanProp}
\end{align}
which involves  the modified Bessel functions $I_\ell(\cdot)$, preserves the normalization of $w(\alpha,m;t)$. In the limit of vanishing diffusion, $D_{\rm rot} \approx 0$, \eqref{eq:PlanTE} describes the classical shearing of the Wigner function, $T_t(\alpha',\ell) \approx \delta(\alpha') \delta_{\ell 0}$.

Angular momentum diffusion broadens the momentum distribution since the energy increases linearly with time, $\partial_t\left<\op{p}_{\alpha}^2/2 I\right>=D_{\rm rot}/I$, which in turn enhances the orientational spread. This is demonstrated in Fig.~\ref{fig:PlanarTE} for the initial superposition state $\psi_0(\alpha) \propto \exp[-\cos^2\alpha/4 \sigma^2_\alpha]$ with Wigner function
\begin{align}
w_0(\alpha,m)=\frac{(-1)^m}{N} I_m \left [ \frac{\cos(2\alpha)}{4 \sigma^2_\alpha} \right ],
\label{eq:OrientGauss}
\end{align} 
where the normalization is $N = 2 \pi I_0 ( 1 / 4 \sigma^2_\alpha)$.

In order to quantify how the orientational spread increases due to the CSL modification, we evaluate the variance $\sigma^2_{\rm C}(t) = 1 - \left \langle \textbf{e}(\upalpha) \right \rangle_t^2$, with $\textbf{e}(\upalpha) = (\cos \upalpha,\sin \upalpha)$ the unit vector in the plane \cite{fischer2014decoherence}. Equation (\ref{eq:PlanTE}) yields
\begin{align}
{\sigma_{\rm C}^2}(t) =& 1-\left(1-\sigma^2_0(t)\right)\exp\left\{-\frac{D_{\rm rot}t}{2\hbar^2}\left[1-{\rm sinc}\left(\frac{2\hbar t}{I}\right)\right]\right\}.
\label{eq:mv}
\end{align}
The unperturbed variance 
\begin{align}
{\sigma_0^2}(t) =& 1-\left[\left<\cos\left(\upalpha+\frac{\hbar\op{m}t}{I}\right)\right>^2_0+\left<\sin\left(\upalpha+\frac{\hbar\op{m}t}{I}\right)\right>^2_0\right]
\label{eq:uv}
\end{align}
is here evaluated with respect to the initial state.

One observes from \eqref{eq:uv} that in the absence of CSL, $D_{\rm rot} = 0$, the initial variance recurs at integer multiples of the revival time $\pi I/\hbar$. Equation~(\ref{eq:mv}) shows that the CSL modification suppresses these revivals on the time scale $2\hbar^2/D_{\rm rot}$. As evident from \eqref{eq:mv}, the CSL modification enhances the orientational spread at all times. 

\section{Conclusion}

The master equation derived in this article shows how the
orientational localization induced by the CSL model leads to a heating of the center-of-mass and the rotational motion. The associated diffusion of the linear and the angular momentum can be conveniently expressed in terms of geometry tensors involving the form factor. For the special case of planar rotations we illustrated how the CSL modification can suppress quantum behavior and lead to an appreciable enhancement of the orientational spread.

Our work clarifies the role of the nonspherical shape of a nanoparticle and shows that strong anisotropies can contribute  substantially to the CSL-induced heating of its motion.
Remarkably, we find that the combined measurement of translational and rotational heating of a single particle would allow one to determine the individual CSL parameters even if the particle is small compared to the CSL localization length.

\subsubsection*{Acknowledgment}
We thank Stefan Nimmrichter for helpful discussions.

\appendix

\begin{widetext}
\section{Diffusion coefficients for Cylinders} 
A calculation of the diffusion coefficients $D_{\perp},D_{\parallel}$ and $D_{\rm rot}$ as defined in Eq.~(\ref{LocParameterCylSymBody}) for a cylinder of length $L$ and radius $R$ yields
\begin{subequations}
\begin{align}
D_{\parallel}=&\frac{\lambda_{\rm C}\hbar^2}{2r_{\rm C}^2}\frac{M^2}{m_0^2R_{\rm C}^2L_{\rm C}^2}h_1(L_{\rm C})\left\{
1-e^{-R_{\rm C}^2}\left[I_0\left(R_{\rm C}^2\right)+I_1\left(R_{\rm C}^2\right)\right]
\right\},\label{StefansErgebnis}\\
D_{\perp}=&\frac{\lambda_{\rm C}\hbar^2}{2r_{\rm C}^2}\frac{M^2}{m_0^2R_{\rm C}^2L_{\rm C}^2}h_2(L_{\rm C}) e^{-R_{\rm C}^2}I_1\left(R_{\rm C}^2\right),\\
D_{\rm rot}=&\frac{\lambda_{\rm C}\hbar^2}{2}\frac{M^2}{m_0^2 R_{\rm C}^2 L_{\rm C}^2}\Bigg(
\frac{R_{\rm C}^2}{2} h_1(L_{\rm C})\left\{ 1 - 2 e^{-R_{\rm C}^2}\left[I_0\left(R_{\rm C}^2\right) + 
\left(1 - \frac{5}{3R_{\rm C}^2}\right)I_1\left(R_{\rm C}^2\right)
\right]\right\}+\frac{L_{\rm C}^2}{3}e^{-R_{\rm C}^2} I_1\left(R_{\rm C}^2\right)\left[
h_2(L_{\rm C})-2
\right]\nonumber\\
&\hspace{9mm}+\left\{
1-e^{-R_{\rm C}^2}\left[I_0\left(R_{\rm C}^2\right)+2I_1\left(R_{\rm C}^2\right)\right]
\right\}\left[h_1(L_{\rm C})-h_2(L_{\rm C})
\right]
\Bigg),
\label{betaPhiCylinder}
\end{align}
\end{subequations}
where $I_0(\cdot),I_1(\cdot)$ denote modified Bessel functions. We abbreviated $R_{\rm C}=R/\sqrt{2}r_{\rm C}$, $L_{\rm C}=L/2r_{\rm C}$,
\begin{equation}
h_1(L_{\rm C}) = 1-e^{-L_{\rm C}^2}, \qquad \text{and} \qquad h_2(L_{\rm C}) =\sqrt{\pi} L_{\rm C}\erf\left(L_{\rm C}\right) - h_1(L_{\rm C}),
\end{equation}
with $\erf(\cdot)$ the error function. In the limit of thin disks, $R / r_{\rm C} \to \infty$, at constant volume $\pi R^2 L$, the rotational diffusion coefficient approaches $D_{\rm rot} \to \lambda_{\rm C}\hbar^2M^2/4m_0^2$, while in the case of long rods we have  $D_{\rm rot} \sim \sqrt{\pi} \lambda_{\rm C}\hbar^2M^2 L/24 m_0^2 r_{\rm C}$ as  $L / r_{\rm C} \to \infty$. Note that the spatial diffusion coefficient along the cylinder's symmetry axis (\ref{StefansErgebnis}) was already derived in Ref. \cite{nimmrichter2014optomechanical}.
\end{widetext}


\end{document}